\documentstyle[12pt]{article}

\title{\begin{flushright}
{\normalsize NUC-MINN-96/15-T\\
December 1996 \\}
\end{flushright}
\vspace*{0.3in}
{\bf COULOMB EFFECTS ON\\ CHARGED KAON DISTRIBUTIONS\\ FROM VLASOV DYNAMICS}}
\author{{\bf Alejandro Ayala}$^{\dagger}$ \\
 {\it Department of Physics}\\
  {\it University of Illinois}\\ \vspace*{0.2in}
 {\it Urbana, IL 61801-3080}\\
{\bf Joseph Kapusta}$^{\ddagger}$ \\
  {\it School of Physics and Astronomy}\\
   {\it University of Minnesota}\\ {\it Minneapolis, MN 55455}}

\date{}

\begin{document}

\maketitle

\begin{center}
Abstract
\end{center}

\noindent
We compute the influence of Coulomb effects on the charged kaon
distribution at low momentum in high energy nucleus-nucleus collisions.
This is accomplished by solving the Vlasov equation for the kaons in
the presence of an expanding, highly charged, fireball.

\vspace{1.in}

\noindent
${}^{\dagger}$ ayala1@rsm1.physics.uiuc.edu \\
\noindent
${}^{\ddagger}$ kapusta@physics.spa.umn.edu

\newpage

The main goal of colliding heavy nuclei at high energy is to produce high
density, high temperature matter, either hadronic or quark--gluon plasma.
One way of studying the properties of high energy density matter
is to compare the single--particle momentum distributions to those
in a nucleon--nucleon collision at the same beam energy.
The many--body medium should be most influential on the low momentum,
or long deBroglie wavelength, part of the spectra and least
influential on the high momentum, or short deBroglie wavelength,
part.  For example, there was a preliminary report of an enhancement
of the charged kaon distributions in the low momentum region
as compared to nucleon--nucleon collisions in an experiment at
Brookhaven National Laboratory's AGS \cite{Barrette}.
Although this enhancement is now known to have been caused by an albedo
source at the spectrometer's collimator edge close to the beam
trajectory \cite{Lacasse}, it begs the question of how the enormous
electrical charge in a central gold--gold collision ($Z = 158$)
modifies the low momentum spectra of charged hadrons.

The most widely used approach to incorporate Coulomb effects
into the description of the charged particle distributions is by
means of the Gamow factor \cite{Gyulassy}.
\begin{equation}
G(\eta) = |\psi(r=0)|^2 = \left| \frac{2\pi\eta}{e^{2\pi\eta}-1} \right|
\label{eq:Gamow}
\end{equation}
Here $\eta= Z_1Z_2\alpha/v$, $v$ is the relative speed of the particle
(charge $Z_1$) with the source (charge $Z_2$),
and $\psi(r=0)$ is the relative Coulomb wave function evaluated
at zero separation between the two (point) charges. In this approach,
the charged particle distribution is given in terms of the neutral
particle distribution and $G$ by
\begin{equation}
\frac{d^3N^{{\rm ch}}(p)}{dp^3} \approx
\frac{d^3N^0(p + k(p))}{dp^3} \times G(\eta)
\label{eq:Gdist}
\end{equation}
where $k(p)$ is the shift in the momentum due to the Coulomb kick.

The original motivation for the introduction of
eq.~(\ref{eq:Gamow}) is the similarity to barrier penetration in fission
and because this formula arises in a perturbative calculation
to first order in $Z\alpha$ for low momenta.
But for heavy ion reactions, where the total charge $Z$ is such that
$Z\alpha \sim 1$, a first order perturbative approach to Coulomb effects is
likely to fail and so will an approach that mimics such.
Figure 1 shows the neutral, positive, and negative kaon momentum
distributions in the fireball rest frame obtained from
eq.~(\ref{eq:Gdist}) using a Maxwell--Boltzmann distribution for
neutrals and a non-expanding fireball with charge $Z=158$ and
kaon temperature $T=120$ MeV.
The neutral distribution is normalized to unity.
Generally the Gamow--modifed charge distributions do not preserve
the normalization so the areas under those curves are not 1.
Figure 2 shows the exact classical result obtained from the formula
\begin{equation}
\frac{dN}{dE} = \int d^3r d^3v f({\bf r},{\bf v},t_0)
\delta(E - mv^2/2 -V_C(r))
\end{equation}
where $V_C(r)$ is the Coulomb potential felt by the kaon.
The initial phase space distribution $f$ is taken to be a
Maxwell--Boltzmann equilibrium distribution
\begin{equation}
   f{\bf r},{\bf v},t_0) \propto
 \exp\{-[mv^2/2 + V_C(r)]/T\} \, \Theta (R_0 - r)
\end{equation}
which constrains the kaons to lie within a uniform density
charged fireball of radius $R_0 = 8$ fm.
The differences between the two figures are quite apparent and indicates
the extent to which the Gamow--modified neutral spectrum can be relied on.

A more important physical effect which the Gamow factor (1)
does not take into account at all is that in a high energy heavy ion
collision the matter blows apart, and in fact it is the low energy
detected charged particle which is left behind.

Therefore it is desirable to look for an alternative approach
to the description of Coulomb effects on charged particle spectra in high
energy heavy ion collisions.  In this paper we explore the influence
of a time--dependent electric field produced by an expanding charged
fireball on the low momentum spectrum of a test charge particle.
To be suitable as a test particle the hadron must not be produced
in great abundance and its nonelectromagnetic interaction with
other hadrons at low relative momentum must not be very strong,
otherwise it will influence the expansion of
the fireball and the problem must be solved self--consistently.
Kaons are good test particles at AGS type energies.
For simplicity and definiteness we assume a thermal momentum
distribution at the strong interaction freeze out and
dynamically evolve this distribution by means of the Vlasov equation.
The interaction of a single test particle with the rest of the matter
is purely electromagnetic.  Since the particles that produce the
electric field are moving away from the collision region, the external
field is time dependent.  We describe the motion of non--relativistic
test particles, thus ignoring interactions with the magnetic component
of the electromagnetic field.  We model the expanding fireball
as a uniformly charged sphere, as viewed from the center-of-mass
of the colliding nuclei, with a radius growing linearly with time.
This expansion model is motivated by the similarity solution to
a hydrodynamically expanding fireball \cite{Laszlo & Joe}.
This model turns out to be solvable analytically albeit in terms
of Bessel functions.  A serious comparison with experiment may
require a simulation based on an event generator like ARC \cite{ARC}
or RQMD \cite{RQMD}.  Finally, we remark that a description in terms
of classical equations of motion for the test charge should be
adequate for kaons with kinetic energy greater than about 10 MeV.
At this energy the kaon DeBroglie wavelength is 2 fm which is small
enough compared to expected gradients of temperature and density in
the fireball.

A uniformly charged sphere which has a radius $R_0$ at time $t_0$,
whose total charge is $Ze$, and whose radius $R$ is increasing with
time at a constant speed $v_s$ produces an electric potential
\begin{equation}
  V(r,t)=\left\{ \begin{array}{ll}
  Ze/4\pi r \,\,,   &   r \geq R = v_st \\
  Ze(3R^2-r^2)/8\pi R^3 \,\,,   &   r \leq R = v_st
  \end{array}
          \right. \, .
   \label{eq:potential}
\end{equation}
The fireball parameters are not independent but are related by
$R_0 = v_s t_0$.  We first concentrate
on the interior $(r\leq v_st)$ potential.
If $f^{\pm}({\bf r},{\bf v},t)$ represents the $\pm e$ charge
particle phase space distribution then, when ignoring particle collisions
after decoupling, its dynamics is governed by Vlasov's equation.
In the interior region
\begin{equation}
   \left[ \frac{\partial}{\partial t} +
   {\bf v}\cdot\frac{\partial}{\partial{\bf r}}
   \pm \frac{t_s}{4t^3}{\bf r}\cdot
   \frac{\partial}{\partial{\bf v}}\right]
   f^{\pm}({\bf r},{\bf v},t)=0\, ,
   \label{eq:Vlasov}
\end{equation}
where
\begin{equation}
t_s \equiv \frac{Ze^2}{\pi mv_s^3} \, ,
   \label{eq:gamma}
\end{equation}
$m$ is the kaon's mass, and $t \geq t_0$. Notice that
$c$, the speed of light, does not enter into the definition of the
{\it characteristic time} $t_s$ since we work in the non--relativistic limit.

The solution to eq.~(\ref{eq:Vlasov}) is found by the method of characteristics.
This involves solving the classical equations of motion and using
the solutions to evolve the initial distribution
$f^{\pm}({\bf r},{\bf v},t_0)$ forward in time.

In the interior the equations of motion are
\begin{eqnarray}
\frac{d{\bf r}}{dt}&=&{\bf v} \, ,\nonumber \\
   \frac{d{\bf v}}{dt}&=&\pm\frac{t_s}{4 t^3}{\bf r} \, ,
   \label{eq:twodyn}
\end{eqnarray}
or equivalently
\begin{equation}
   \frac{d^2{\bf r}}{dt^2}=\pm \frac{t_s}{4t^3}{\bf r} \, .
   \label{eq:onedyn}
\end{equation}
For the initial distribution we take a Maxwell--Boltzmann equilibrium
distribution at temperature $T$.
\begin{equation}
   f^{\pm}({\bf r},{\bf v},t_0) \propto
 \exp\{-[mv^2/2 \pm e V(r,t_0)]/T\} \, \Theta (R_0 - r)
   \label{eq:initial}
\end{equation}
Note that this constrains the kaons to lie within the initial sphere.

The solution to the equations of motion for positive kaons is
\begin{equation}
{\bf r}(t) = {\bf A}\sqrt{t}I_1\left(\sqrt{t_s/t}\right)
+ {\bf B}\sqrt{t}K_1\left(\sqrt{t_s/t}\right)
\end{equation}
and for negative kaons is
\begin{equation}
{\bf r}(t) = {\bf A}\sqrt{t}J_1\left(\sqrt{t_s/t}\right)
+ {\bf B}\sqrt{t}N_1\left(\sqrt{t_s/t}\right)
\end{equation}
where the constant vectors ${\bf A}$ and ${\bf B}$ are determined by
the initial conditions.
The solution to eq.~(\ref{eq:Vlasov}) incorporating the initial condition
is
\begin{eqnarray}
   f^{\pm}({\bf r},{\bf v},t)&\!\!\!=\!\!\!&{\cal N}
   \exp\left\{-\frac{m}{2T}\left(
    \left[\frac{c_1(t) {\bf r}}{t_s} - c_2(t) {\bf v}\right]^2
    \mp \frac{t_s^3}{4t_0^3}
    \left[\frac{c_3(t){\bf r}}{t_s}-c_4(t) {\bf v}\right]^2
    \right) \right\} \nonumber \\
    &\!\!\!\times\!\!\!& \Theta(R_0-|c_3(t){\bf r}-c_4(t)t_s{\bf v}|))\, .
       \label{eq:solution}
\end{eqnarray}
The normalization constant ${\cal N}$ is chosen so
that the distribution is normalized to unity at the initial time $t_0$.
\begin{equation}
\int d^3r\,d^3v\,f^{\pm}({\bf r},{\bf v},t_0) = 1
\end{equation}
Of course, this ensures that the distribution is normalized to unity
for all later times, even when account is taken of the fact that
the solution differs outside the expanding sphere.

For negative kaons the dimensionless functions of time $c_i(t)$ are given by
\begin{eqnarray}
\frac{c_1(t)}{t_s}&=&\frac{\pi}{4} \frac{t_s}{t_0 t}
 \left[ J_2\left(\sqrt{t_s/t_0}\right) N_2\left(\sqrt{t_s/t}\right) -
  N_2\left(\sqrt{t_s/t_0}\right) J_2\left(\sqrt{t_s/t}\right) \right] \nonumber \\
c_2(t)&=&\frac{\pi}{2} \sqrt{\frac{t_st}{t_0^2}}
 \left[ J_2\left(\sqrt{t_s/t_0}\right) N_1\left(\sqrt{t_s/t}\right) -
  N_2\left(\sqrt{t_s/t_0}\right) J_1\left(\sqrt{t_s/t}\right) \right]
  \nonumber \\
\frac{c_3(t)}{t_s}&=&\frac{\pi}{2} \sqrt{\frac{t_0}{t_st^2}}
 \left[ J_1\left(\sqrt{t_s/t_0}\right) N_2\left(\sqrt{t_s/t}\right) -
  N_1\left(\sqrt{t_s/t_0}\right) J_2\left(\sqrt{t_s/t}\right) \right]
  \nonumber \\
c_4(t)&=&\pi \sqrt{\frac{t_0 t}{t_s^2}}
 \left[ J_1\left(\sqrt{t_s/t_0}\right) N_1\left(\sqrt{t_s/t}\right) -
  N_1\left(\sqrt{t_s/t_0}\right) J_1\left(\sqrt{t_s/t}\right) \right] \, .
   \label{eq:taus}
\end{eqnarray}
For positive kaons the corresponding $c_i(t)$ are obtained
by the substitution $t_s \rightarrow -t_s$ in the right--hand side of 
eqs.~(\ref{eq:taus}) and the identities
\begin{eqnarray}
N_n(ix) &=& i^{n+1}I_n(x)
- \frac{2}{\pi} (-i)^n K_n(x) \, ,\nonumber \\
J_n(ix) &=& i^nI_n(x)
   \label{eq:substitution}\, .
\end{eqnarray}

Generally the asymptotic momentum distribution is obtained by
integrating over position and taking the limit $t\rightarrow\infty$.
Once a kaon has crossed the surface of the expanding sphere
it will see a time--independent potential, hence its energy
will thereafter be conserved.  Since a positive kaon will see
a positive potential, its kinetic energy will not decrease, and neither
will its velocity.  Once a positive kaon has crossed the surface
it will not come back.  This will not be true for negative kaons;
they will be discussed later.
Due to spherical symmetry, the distribution is a function only
of the magnitude of the three dimensional momentum vector.
The coordinate integration can be performed analytically
with the result for positive kaons being
\begin{displaymath}
\hspace*{-3in}
\frac{d^3N}{d^3p}=\pi {\cal N} \left(\frac{R_0}{\chi \delta}\right)^3
 {\rm exp}\!\left(-p^2/2mT\chi^2\right)
\end{displaymath}
\begin{equation}
\times \left\{ \frac{\sqrt{\pi}}{2i}
\left[ \Phi\left(\frac{p + \delta p_C}{ip_C}\right)\! - \!
 \Phi\left(\frac{p - \delta p_C}{ip_C}\right) \right]
 {\rm exp}\left(-p^2/p_C^2\right)
+ {\rm exp}\!\left(\delta^2\right) \frac{{\rm sinh}(2 \delta p /p_C)}{p/p_C}
\right\}
   \label{eq:def1}
\end{equation}
where $\Phi(x)$ is the error function and
\begin{eqnarray}
p_C^2 &=& \frac{8mTt_0}{t_s}  \left( I_2^2- I_1^2 \right)
 \frac{I_1^2}{I_2^2} \nonumber \\
\chi^2 &=& \frac{4t_0}{t_s} I_1^2 \nonumber \\
\delta^2 &=& \frac{m v_s^2 t_s}{8T t_0}
\left(\frac{I_2^2}{I_1^2} - 1 \right)
   \label{eq:def2}
\end{eqnarray}
where the Bessel functions are all evaluated at $\sqrt{t_s/t_0}$.
Taking $Z$ = 158, $R_0$ = 8 fm, $v_s/c$ = 0.4, and $T$ = 120 MeV,
results in $t_0$ = 20 fm/c, $t_s$ = 28.7 fm/c, $p_C$ = 1390 MeV/c,
$\chi$ = 1.19, and $\delta$ = 0.330.
It is important to remember that eq.~(\ref{eq:def1}) is valid only
so long as the test particle (positive kaon) remains within the
expanding sphere.  This will be the case if its asymptotic speed
is less than $v_s$.

To compute the distribution of positive kaons for asymptotic speeds
greater than $v_s$ is straightforward due to the fact that
once it has crossed the boundary of the expanding sphere, it sees a
time independent potential so that its energy is thereafter conserved.
The flux of particles crossing the surface at time $t$ is
\begin{displaymath}
\int d^3v \,{\bf v} \cdot \hat{{\bf r}} f^+({\bf r},{\bf v},t)|_{r = v_st}
\end{displaymath}
which is to be evaluated at $r = v_s t$.  The energy distribution is
obtained by multiplying by the surface area, inserting an energy
conserving $\delta$ function, and integrating over time.
\begin{eqnarray}
dN^+/dE &=& \int_{t_0}^{\infty}dt \, v_s^2 t^2 \int d\Omega_{\bf r}
\int d^3v \,{\bf v} \cdot \hat{\bf {r}} f^+({\bf r},{\bf v},t)
|_{r = v_st} \nonumber \\
&\times& \delta(E - p^2/2m - Ze^2/4\pi v_st)
\label{eq:flux}
\end{eqnarray}
The asymptotic momentum distribution is readily obtained from the
energy distribution.

Figure 3 shows the $K^+$ distributions for expanding fireballs
with initial radius 8 fm, kaon temperature 120 MeV, and expansion
velocities $v_s/c$ = 0.2, 0.4 and 0.6.  For comparison the dotted curve
shows the asymptotic distribution obtained from a static fireball.
There is a discontinuity in the slope of the distributions at
$E = \frac{1}{2}mv^2_s$.  We have checked that the areas under all
curves are 1.  The significant physics here is that the distribution
remains finite at zero energy.  The reason is that the charged
fireball blows apart and leaves a finite number of low energy kaons
behind.  This cannot happen with a static fireball where a classical
turning point exists.  In the limit that the fireball expansion
velocity is large compared to the thermal velocity of the kaons,
the kaon distribution is hardly changed from the pure exponential
form.  The fireball expands so fast that it leaves the kaon
distribution frozen to its initial shape.  In the limit that the
expansion velocity goes to zero, the static fireball result is
recovered.

Now we turn to negative kaons.  A negatively charged kaon can cross
the fireball's surface if it has a sufficiently large radial velocity.
After crossing the surface, it will slow down due to the attractive
potential.  It may even slow enough so that the surface passes it by!
So the possibilities are three--fold: it always remains within the
expanding sphere, it crosses the surface once and never comes back, or
it crosses the surface twice so that asymptotically it remains within
the expanding sphere.  Kaons with an asymptotic energy greater than
$\frac{1}{2}mv_s^2$ can be dealt with by the flux formula
(\ref{eq:flux}).  Kaons with an asymptotic energy less than this
cannot be handled analytically.  This is ultimately due to the fact
that the Coulomb trajectories cannot be expressed in terms of
elementary functions of the time and so it is not possible to
construct the phase space distribution outside the expanding
sphere in a useful form.  To handle this situation we have written
some computer programs which calculate the trajectory of a kaon
given its initial phase space position.  One program was written
in turbopascal with a screen display in real time which graphically
shows the behaviors sketched above.  A second program was written
in fortran which connects the initial phase space point to the final
energy (usually the asymptotic energy was determined at 100$t_0$).

Figure 4 shows the $K^-$ distributions corresponding to the
same fireball configurations as in figure 3.  The solid curves
show the result of the flux calculation for asymptotic energies
greater than $\frac{1}{2}mv_s^2$.  The histograms are the result
of numerically computing the kaon trajectories for 720,000 initial
phase space points.  The initial radial position was incremented
in 0.1 fm bins, the initial speed was incremented in 0.01c bins,
and the angle between the initial position and velocity vectors
was incremented in 2$^o$ bins.  (Hence $80\times 100\times 90
= 720,000$.)  The distribution is enhanced at small energies
relative to the neutral kaons and to the positive kaons, but not
as much as for a static fireball, shown in the figure by dotted
curves.  Another distinction is that there are no classically
bound $K^-$, unlike the case of a static fireball.

The kaon data which originally attracted attention was
for rather forward rapidities of 2.1 to 2.5 in contrast to
the rapidity of the nucleus-nucleus center-of-mass frame
of about 1.5.  Plotted in the nucleus-nucleus center-of-mass
frame the kinetic energies go no lower than 80 MeV for $K^-$
and 70 MeV for $K^+$.  It is problematic to compare against
theoretically calculated spectra for kinetic energies greater
than about 100 MeV when a nonrelativistic approximation has
been made.  In addition, the selection criterion was for the
10\% most central collisions, corresponding to the 30\%
most central impact parameters.  These impact parameters are
really too large to focus on truly central collisions of equal
mass nuclei.

In conclusion we have shown the importance of the Coulomb force
on charged kaons in central collisions of large nuclei at high
energies.  We have shown how the expansion of the fireball affects
the kaon distributions in a quantitative way.  Comparison to data
at kinetic energies greater than about 100 MeV probably requires
a relativistic treatment which we are now working on.  By varying
the initial phase space distribution of the kaons by invoking
some component of flow, for example, and comparing with data should
yield quantitative information on the dynamics of kaons and
heavy ion collisions.

\section*{Acknowledgements}

The authors wish to thank J. Stachel and R. Lacasse for discussing
the data of E814/E877. J. K. thanks B. Bayman for a turbopascal
program which solves Newton's equation for the motion of a planet.
A. A. thanks the Institute for Nuclear Theory at the University of
Washington for its hospitality and the Department of Energy for
its support there during the completion of this work.
This work was supported by the US National Science Foundation
under grant NSF PHY94-21309 and by the US Department of Energy
under grant DE-FG02-87ER40328.

\newpage

\section*{Figure Captions}

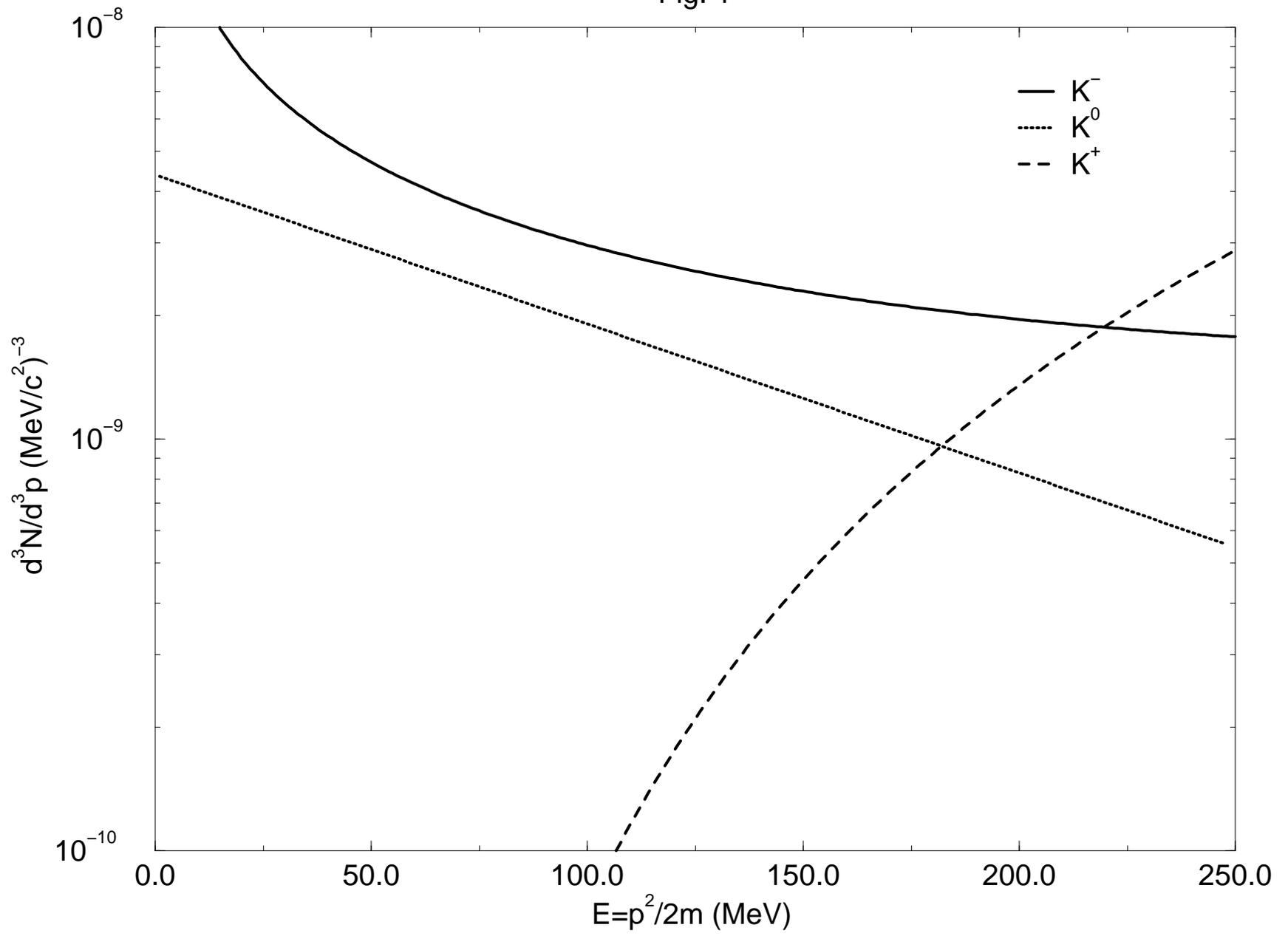
\noindent Fig. 1: Gamow factor momentum distributions $d^3N/d^3p$ as a function
of kinetic energy $E$ for a non--expanding fireball with $T=120$ MeV and
$Z = 158$. Also shown is the corresponding neutral distribution.

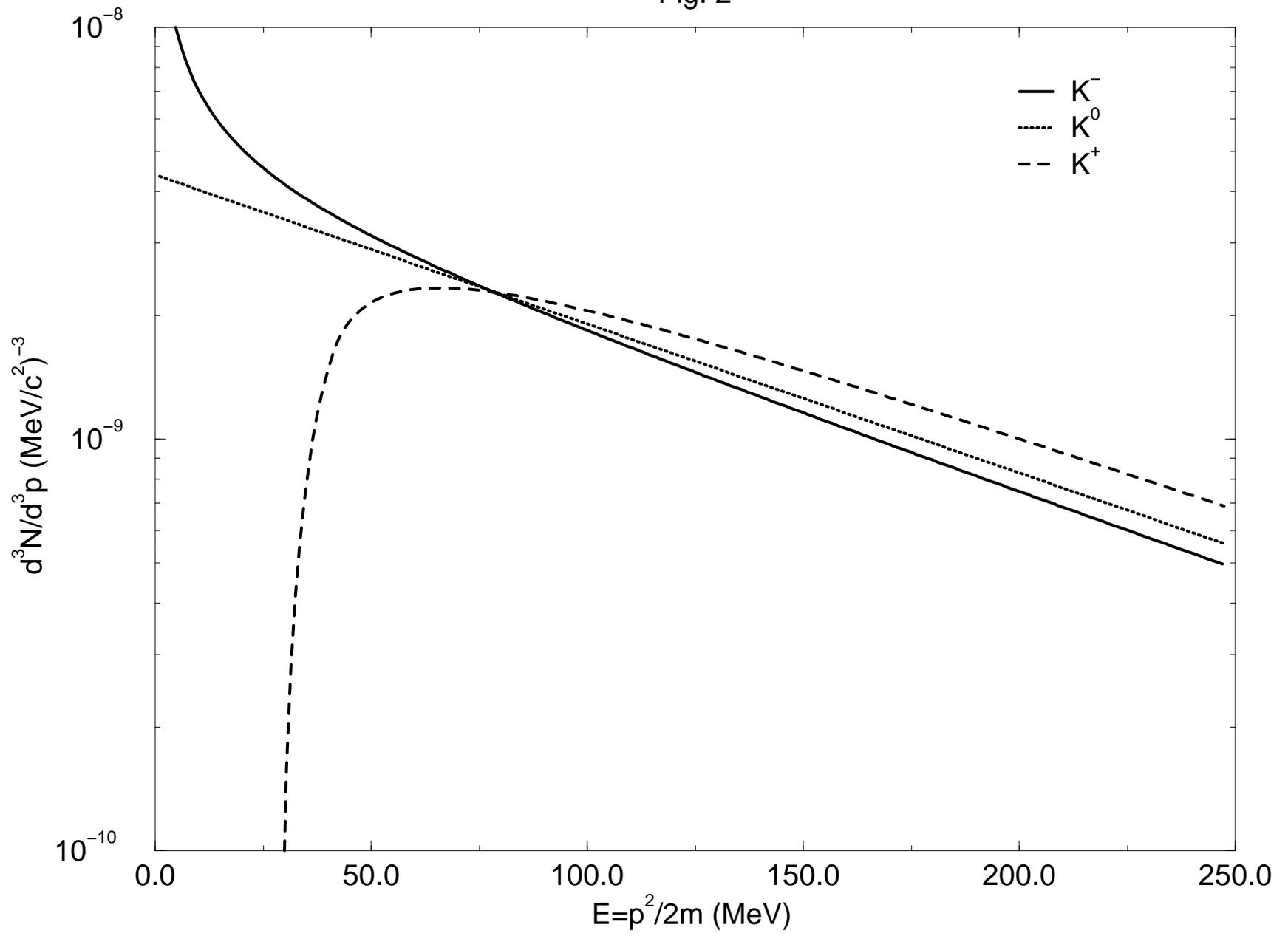
\noindent Fig. 2: Momentum distributions $d^3N/d^3p$ as a function of kinetic 
energy $E$ for a non--expanding fireball with $R_0=8$ fm, $T=120$ MeV, $Z=158$.
The sudden drop in the $K^+$ distribution signals the proximity of
$E$ to $E_{{\rm min}}\equiv Z e^2 /4\pi R_0$, the classical turning point.
Compare this to figure 1.

\noindent Fig. 3: $K^+$ momentum distributions $d^3N/d^3p$ as a function of
kinetic energy $E$ for a uniformly expanding fireball with $R_0=8$ fm,
$T=120$ MeV, $Z=158$ and various values of the expansion velocity
$v_s$ indicated with solid lines.  Also shown is the corresponding
result for a static fireball $v_s=0$ indicated with dotted lines.

\noindent Fig. 4: $K^-$ momentum distributions $d^3N/d^3p$ as a function of
kinetic energy $E$ for a uniformly expanding fireball with $R_0=8$ fm,
$T=120$ MeV, $Z=158$ and various values of the expansion velocity
$v_s$.  The solid lines starting at $E_s = \frac{1}{2}mv_s^2$ are
obtained from the flux across the fireball's surface.  The histograms
are obtained by solving the classical equations of motion numerically
for 720,000 starting points in phase space.  Also shown is the corresponding
result for a static fireball $v_s=0$ indicated with dotted lines.

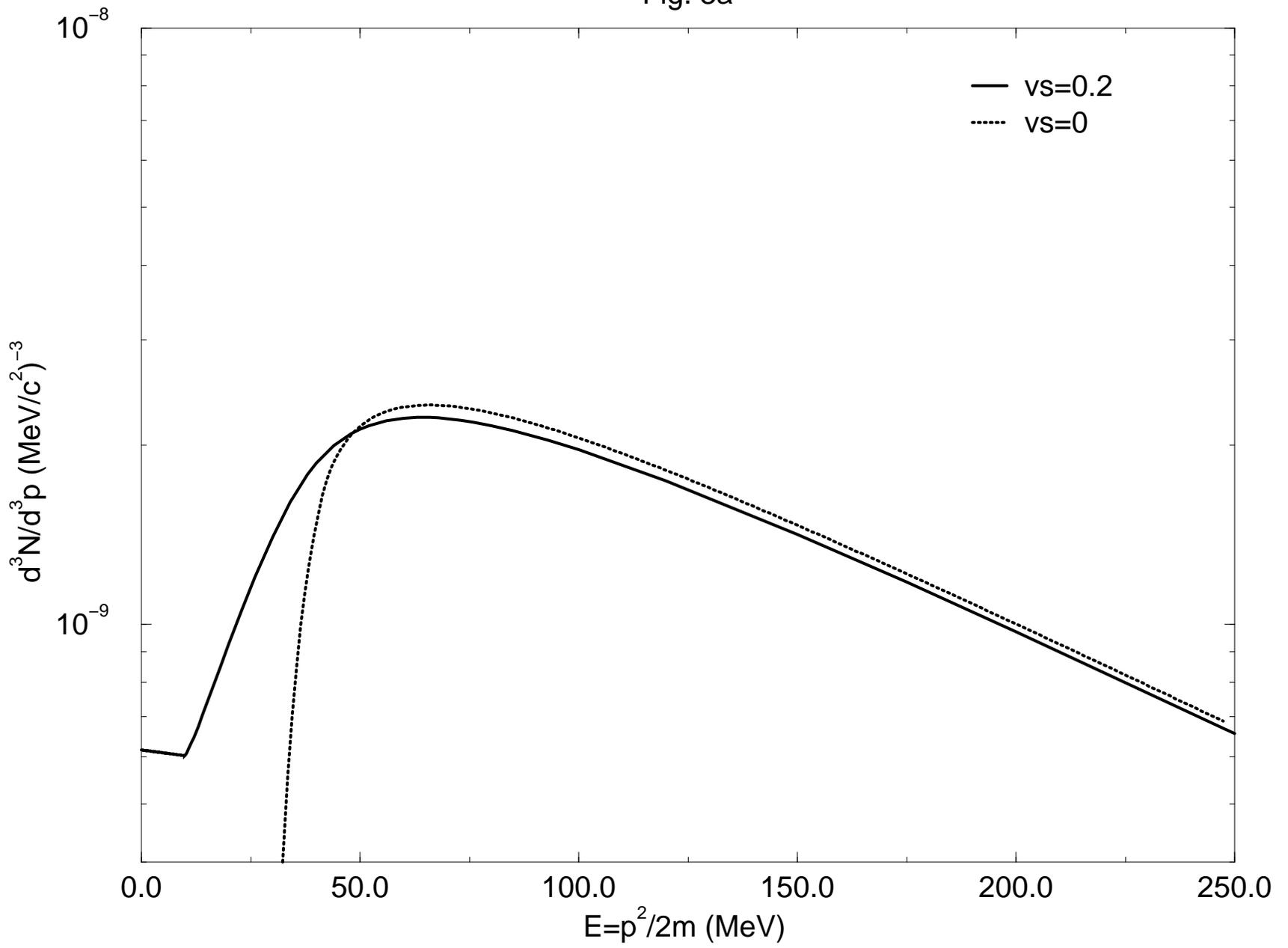
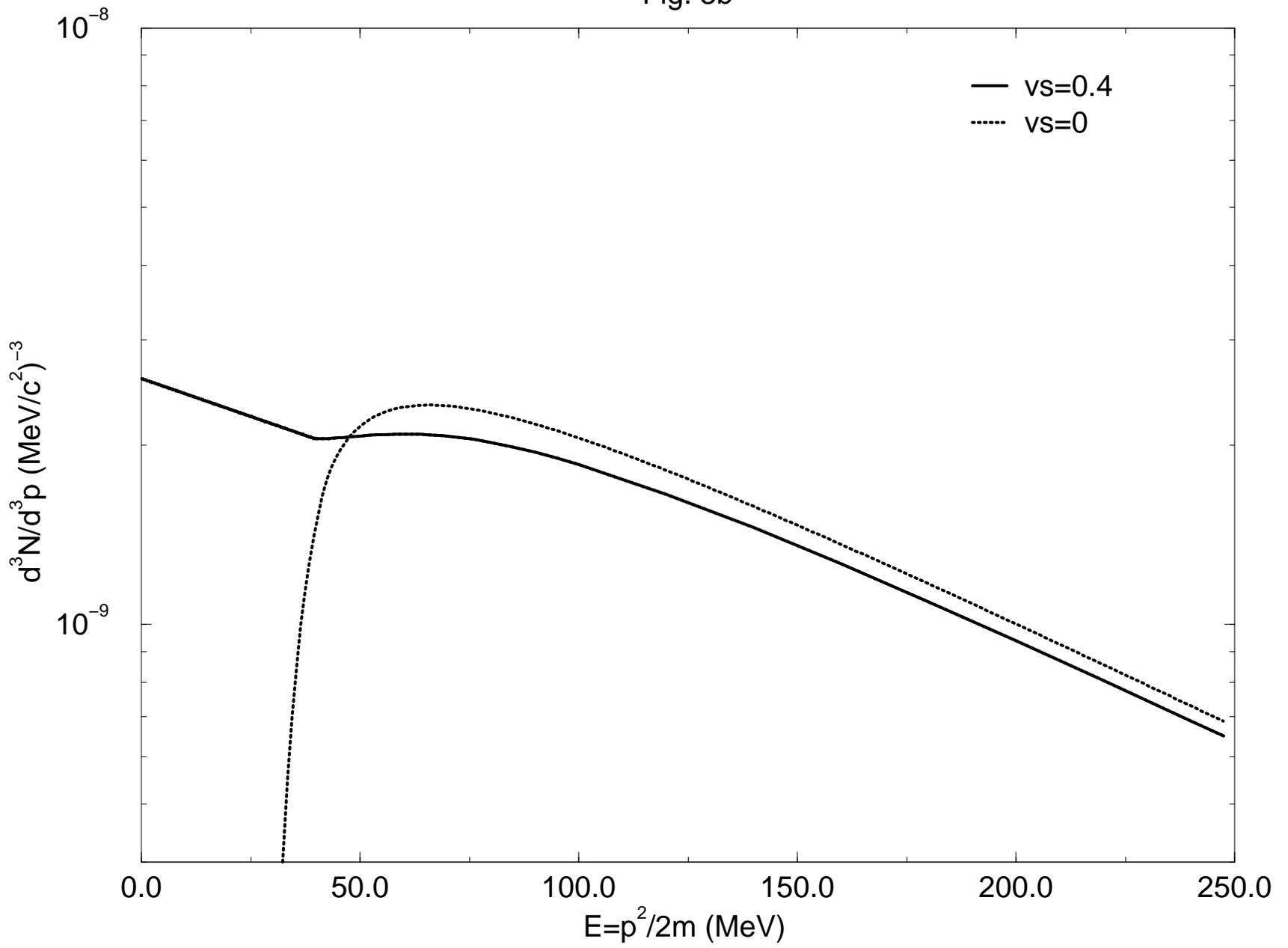
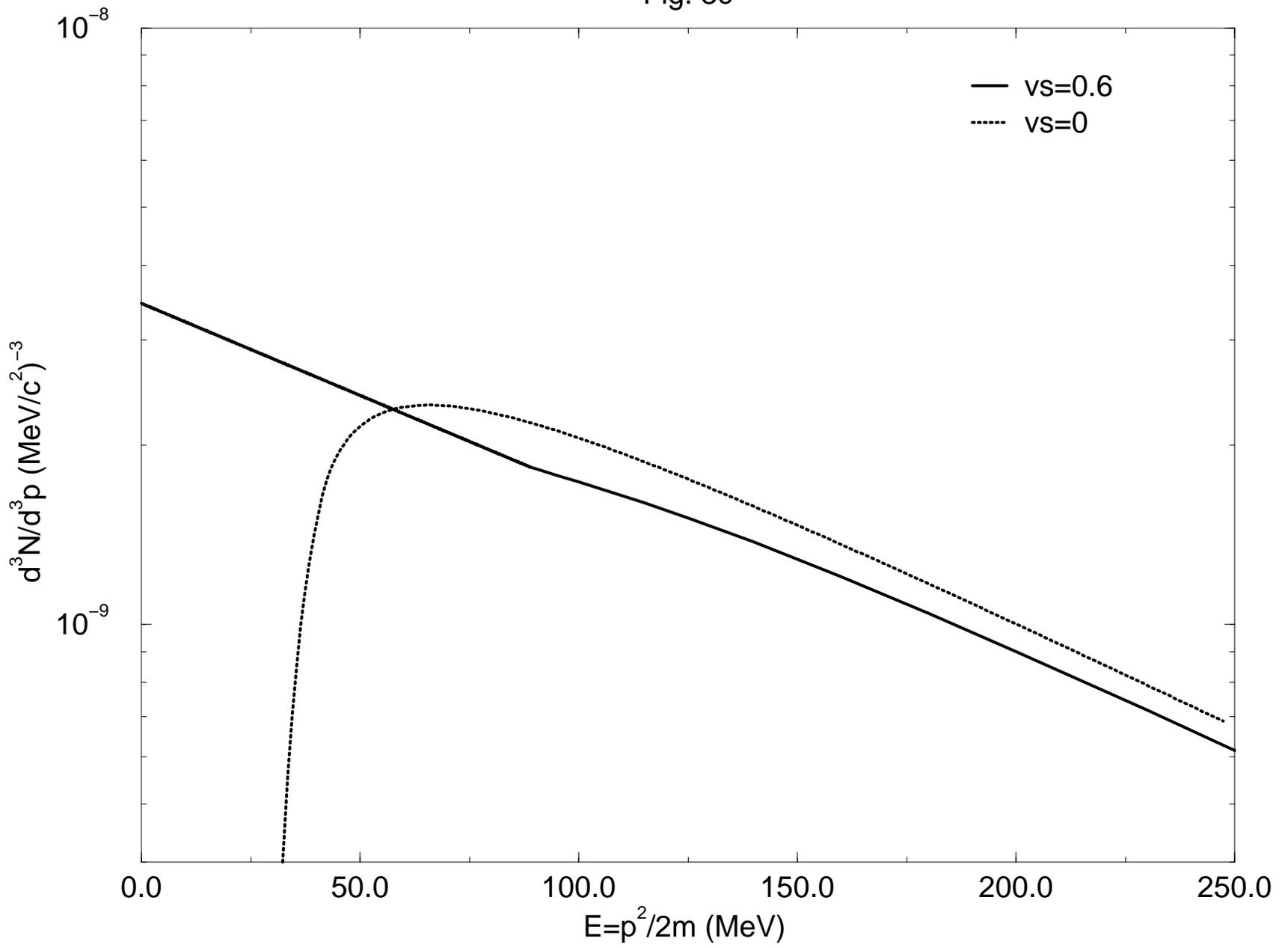
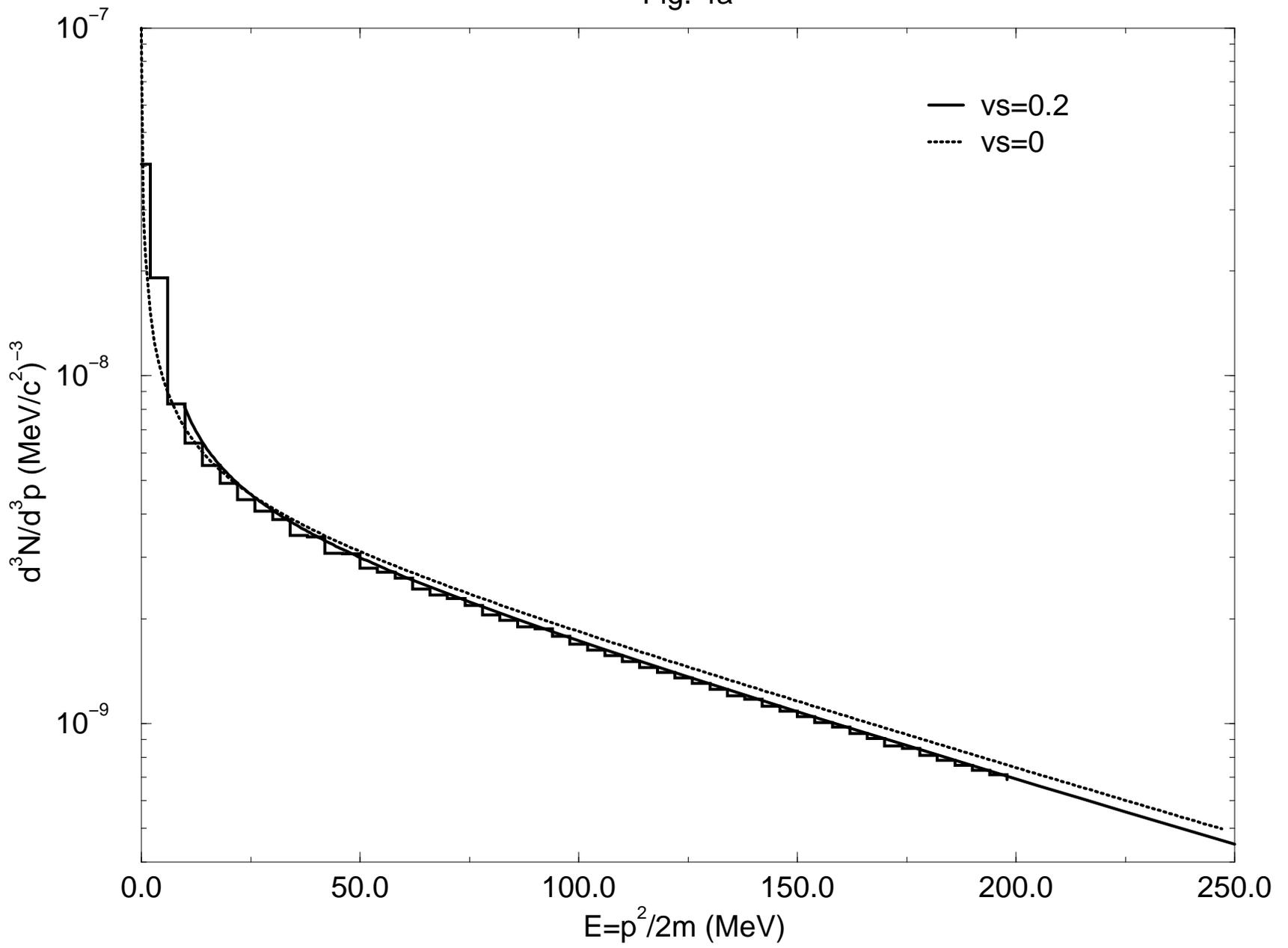
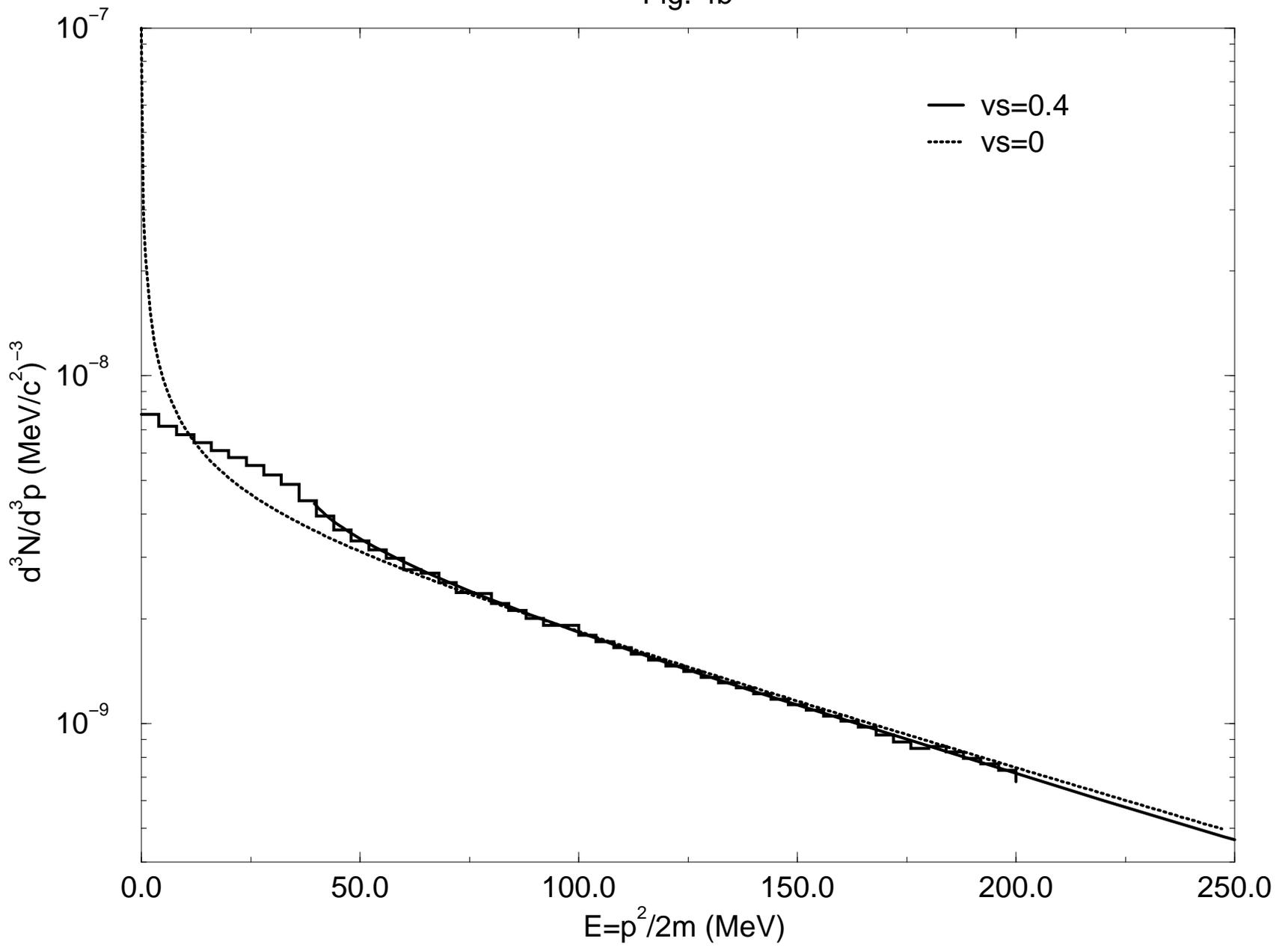
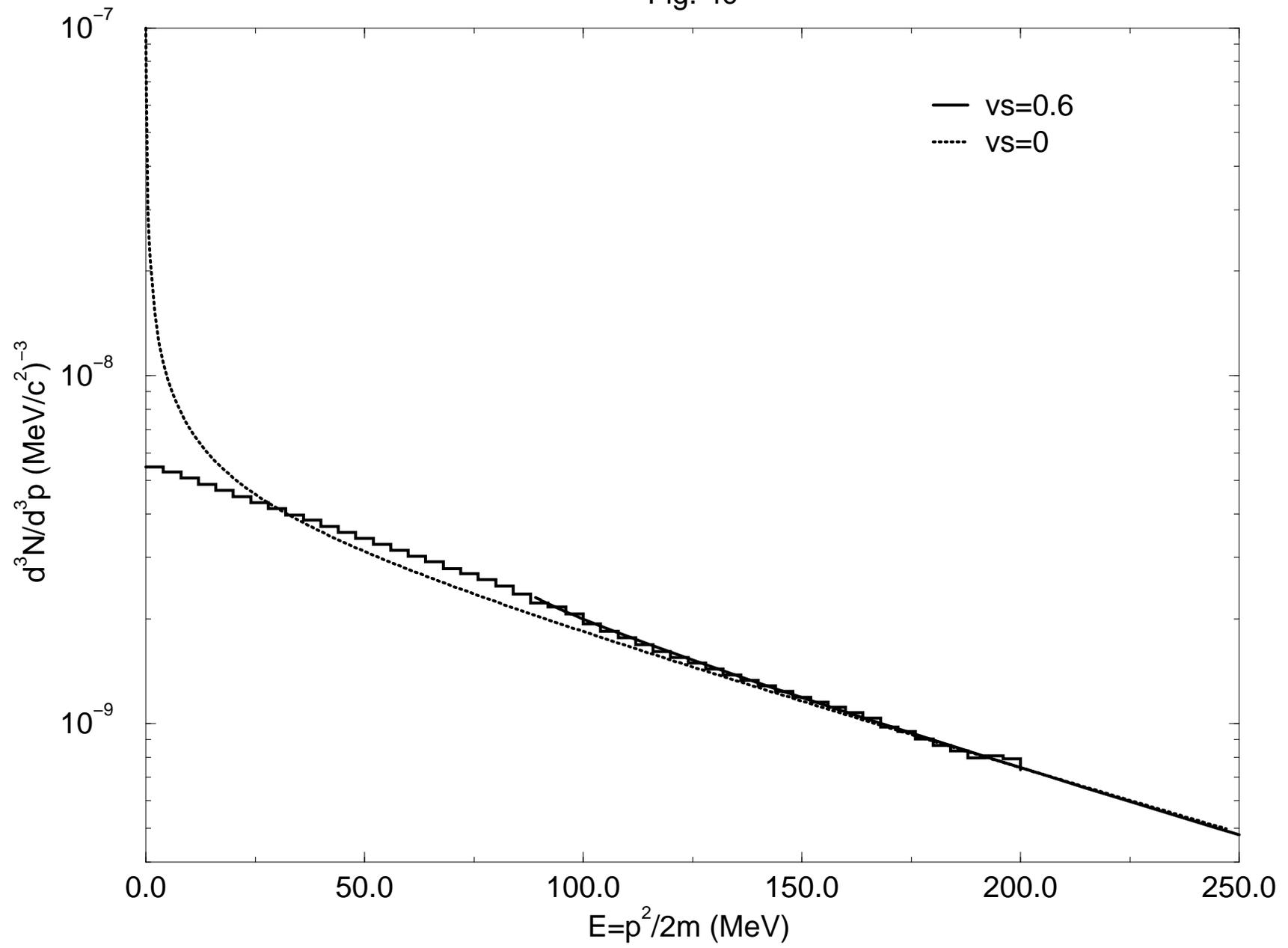
\end{document}